# Measurements of the Proton and Deuteron Spin Structure Function $g_2$ and Asymmetry $A_2$[*]

The E143 Collaboration

K. Abe,[15] T. Akagi,[12,15] P. L. Anthony,[12] R. Antonov,[11] R. G. Arnold,[1] T. Averett,[16,‡‡] H. R. Band,[17] J. M. Bauer,[7] H. Borel,[5] P. E. Bosted,[1] V. Breton,[3] J. Button-Shafer,[7] J. P. Chen,[16] T. E. Chupp,[8] J. Clendenin,[12] C. Comptour,[3] K. P. Coulter,[8] G. Court,[12,*] D. Crabb,[16] M. Daoudi,[12] D. Day,[16] F. S. Dietrich,[6] J. Dunne,[1] H. Dutz,[12,**] R. Erbacher,[12,13] J. Fellbaum,[1] A. Feltham,[2] H. Fonvieille,[3] E. Frlez,[16] D. Garvey,[9] R. Gearhart,[12] J. Gomez,[4] P. Grenier,[5] K. A. Griffioen,[11,†] S. Hoibraten,[16,§] E. W. Hughes,[12,‡‡] C. Hyde-Wright,[10] J. R. Johnson,[17] D. Kawall,[13] A. Klein,[10] S. E. Kuhn,[10] M. Kuriki,[15] R. Lindgren,[16] T. J. Liu,[16] R. M. Lombard-Nelsen,[5] J. Marroncle,[5] T. Maruyama,[12] X. K. Maruyama,[9] J. McCarthy,[16] W. Meyer,[12,**] Z.-E. Meziani,[13,14] R. Minehart,[16] J. Mitchell,[4] J. Morgenstern,[5] G. G. Petratos,[12,‡] R. Pitthan,[12] D. Pocanic,[16] C. Prescott,[12] R. Prepost,[17] P. Raines,[11] B. Raue,[10] D. Reyna,[1] A. Rijllart,[12,††] Y. Roblin,[3] L. S. Rochester,[12] S. E. Rock,[1] O. A. Rondon,[16] I. Sick,[2] L. C. Smith,[16] T. B. Smith,[8] M. Spengos,[1] F. Staley,[5] P. Steiner,[2] S. St.Lorant,[12] L. M. Stuart,[12] F. Suekane,[15] Z. M. Szalata,[1] H. Tang,[12] Y. Terrien,[5] T. Usher,[12] D. Walz,[12] J. L. White,[1] K. Witte,[12] C. C. Young,[12] B. Youngman,[12] H. Yuta,[15] G. Zapalac,[17] B. Zihlmann,[2] D. Zimmermann[16]

[1] *The American University, Washington, D.C. 20016*
[2] *Institut für Physik der Universität Basel, CH–4056 Basel, Switzerland*
[3] *LPC IN2P3/CNRS, University Blaise Pascal, F–63170 Aubiere Cedex, France*
[4] *CEBAF, Newport News, Virginia 23606*
[5] *DAPNIA-Service de Physique Nucleaire Centre d'Etudes de Saclay, F–91191 Gif/Yvette, France*
[6] *Lawrence Livermore National Laboratory, Livermore, California 94550*
[7] *University of Massachusetts, Amherst, Massachusetts 01003*
[8] *University of Michigan, Ann Arbor, Michigan 48109*
[9] *Naval Postgraduate School, Monterey, California 93943*
[10] *Old Dominion University, Norfolk, Virginia 23529*
[11] *University of Pennsylvania, Philadelphia, Pennsylvania 19104*
[12] *Stanford Linear Accelerator Center, Stanford, California 94309*
[13] *Stanford University, Stanford, California 94305*
[14] *Temple University, Philadelphia, Pennsylvania 19122*
[15] *Tohoku University, Sendai 980, Japan*
[16] *University of Virginia, Charlottesville, Virginia 22901*
[17] *University of Wisconsin, Madison, Wisconsin 53706*

We have measured proton and deuteron virtual photon-nucleon asymmetries $A_2^p$ and $A_2^d$ and structure functions $g_2^p$ and $g_2^d$ over the range $0.03 < x < 0.8$ and $1.3 < Q^2 < 10$ (GeV/c)$^2$ by inelastically scattering polarized electrons off polarized ammonia targets. Results for $A_2$ are significantly smaller than the positivity limit $\sqrt{R}$ for both targets. Within experimental precision the $g_2$ data are well-described by the twist-2 contribution, $g_2^{WW}$. Twist-3 matrix elements have been extracted and are compared to theoretical predictions.

Typeset using REVTeX

[*]Work supported in part by Department of Energy contract DE-AC03-76SF00515.

The nucleon spin structure functions $g_1(x,Q^2)$ and $g_2(x,Q^2)$ have in recent times received much theoretical and experimental attention. Experiments at CERN [1–4] and SLAC [5–7] have measured $g_1$ and $g_2$ using deep inelastic scattering (DIS) of longitudinally polarized leptons on polarized nuclear targets. These studies have largely concentrated on $g_1^p$, $g_1^d$, and $g_1^n$, which are dominant when the target is polarized along the beam direction. Their results have established that the quark component of the nucleon helicity is much smaller than the predictions of the naive quark-parton model [8]. In addition, the Bjorken Sum rule [9], a fundamental QCD prediction for the difference of the first moments of $g_1^p$ and $g_1^n$, has been confirmed within the uncertainties of the experiment and theory [3,5,7].

The present work is devoted to $g_2^p(x,Q^2)$ and $g_2^d(x,Q^2)$ which are dominant when longitudinally polarized leptons scatter from transversely polarized nucleons. The $g_2$ spin structure function probes both transverse and longitudinal parton polarization distributions inside the nucleon. Properties of $g_2$ have been established using the operator product expansion (OPE) within QCD [10,11], and the interpretation of $g_2$ in the light-cone parton model is also on firm grounds [12–14]. Three components contribute to $g_2$: a leading twist-2 part, $g_2^{WW}(x,Q^2)$, coming from the same set of operators that contribute to $g_1$, another twist-2 part coming from the quark transverse polarization distribution $h_T(x,Q^2)$, and a twist-3 part coming from quark-gluon interactions $\xi(x,Q^2)$,

$$g_2(x,Q^2) = g_2^{WW}(x,Q^2) - \int_x^1 \frac{\partial}{\partial y}\left(\frac{m}{M} h_T(y,Q^2) + \xi(y,Q^2)\right)\frac{dy}{y}. \qquad (1)$$

The Bjorken scaling variable is denoted by $x$, $Q^2$ is the four-momentum transfer squared, and $m$ and $M$ are quark and nucleon masses. The $g_2^{WW}$ expression of Wandzura-Wilczek [15]

$$g_2^{WW}(x,Q^2) = -g_1(x,Q^2) + \int_x^1 \frac{g_1(y,Q^2)}{y} dy. \qquad (2)$$

can be derived from the OPE [10,11] sum rules for $g_1$ and $g_2$ at fixed $Q^2$

$$\int_0^1 x^n g_1(x,Q^2) dx = \frac{a_n}{2}, \qquad n = 0, 2, 4, ...$$
$$\int_0^1 x^n g_2(x,Q^2) dx = \frac{1}{2}\frac{n}{n+1}(d_n - a_n), \quad n = 2, 4, ... \qquad (3)$$

by keeping $a_n$ (twist-2) and neglecting the $d_n$ (twist-3) matrix elements of the renormalized operators. The quantity $h_T(x,Q^2)$ appearing in Eq. (1) contributes to leading order in quark-quark scattering (e.g., polarized Drell-Yan processes), but is suppressed by $m/M$ [13,14,16] in DIS. This component should not be confused with the twist-3 quark mass term that appears in the OPE, nor with the related average transverse spin [16,17] $g_T = g_1 + g_2$ that measures the spin distribution normal to the virtual photon momentum.

The OPE analysis does not yield a sum rule for the first moment of $g_2$ ($n = 0$). However, Burkhardt and Cottingham [18] have derived the sum rule $\int_0^1 g_2(x) dx = 0$ in the $Q^2 \to \infty$ limit from virtual Compton scattering dispersion relations. Due to the uncertainty in the very small $x$ behavior of $g_2$, it may not be possible to experimentally test this sum rule [10,19].

The spin asymmetries $A_1$, $A_2$ for virtual Compton scattering are directly related to the spin structure functions. From the virtual photon transverse cross section $\sigma_T$ and the transverse-longitudinal interference cross section $\sigma^{TL}$ one can form the transverse asymmetry

$$A_2(x,Q^2) = \frac{\sigma^{TL}}{\sigma^T} = \frac{(Q/\nu)[g_1(x,Q^2) + g_2(x,Q^2)]}{F_1(x,Q^2)} \qquad (4)$$

where $\nu = E - E'$ is the energy transfer, $E$ and $E'$ are the incident beam and scattered lepton energies, and $F_1(x,Q^2)$ is one of the spin-averaged DIS structure functions. The SMC has measured $A_2^p$ [4] (See Fig. 1) at four values of $x$ in the range $0.006 \leq x \leq 0.6$ and $1 < Q^2 < 30$ (GeV/c)$^2$. These results do not violate the positivity condition $|A_2(x,Q^2)| \leq \sqrt{R(x,Q^2)}$, where $R(x,Q^2) = \sigma^L/\sigma^T$ is the ratio of the longitudinal to transverse virtual photon absorption cross sections.

In this paper, we report on measurements of the proton and deuteron asymmetries $A_2^p$ and $A_2^d$ and the transverse structure functions $g_2^p$ and $g_2^d$ from SLAC experiment E143. Results for $g_1^p$ and $g_1^d$ from this experiment as well as details on the experiment and data analysis have been previously reported [6,7]. Longitudinally polarized electrons with energy 29.1 GeV were scattered from polarized protons and deuterons into two independent spectrometers at angles of 4.5° and 7°. The beam polarization, typically $P_b = 0.85 \pm 0.02$, was measured with a Møller polarimeter. The target cells were filled with granules of either $^{15}NH_3$ or $^{15}ND_3$, and were polarized using the technique of dynamic nuclear polarization. The targets could be polarized longitudinally or transversely relative to the beam by physically rotating the polarizing magnet. Target polarization $P_t$, measured by a calibrated NMR, averaged around $0.65 \pm 0.017$ for protons and $0.25 \pm 0.011$ for deuterons.

The experimental asymmetries for transverse ($A_\perp$) and longitudinal ($A_\parallel$) target polarizations were determined from

$$A_\perp \text{ (or } A_\parallel) = C_1 \left( \frac{N_L - N_R}{N_L + N_R} \frac{1}{f P_b P_t} - C_2 \right) + A_{RC}, \tag{5}$$

where $N_L$ and $N_R$ are the number of scattered electrons per incident electron for negative and positive beam helicity, where corrections have been made for charge-symmetric backgrounds and deadtime; $f$ is the dilution factor representing the fraction of measured events originating from polarizable protons or deuterons within the target; $C_1$ and $C_2$ correct for the polarized nitrogen nuclei and for residual polarized protons in the $ND_3$ target; and $A_{RC}$ are the radiative corrections, which include internal [20] and external [21] contributions. These $x$-dependent radiative corrections typically shifted $A_2$ by +0.01. The corresponding shift in $g_2$ was +0.30 at low $x$, decreasing rapidly to +0.002 at high $x$. The systematic errors for the radiative corrections to $A_\perp$ were typically as large as the corrections themselves and were dominated by the uncertainty in the model for $g_2(x, Q^2)$.

Both $A_2$ and $g_2$ can be expressed in terms of the experimental asymmetries as:

$$A_2(x, Q^2) = \frac{\gamma(2-y)}{2d} \left[ A_\perp \frac{y(1+xM/E)}{(1-y)\sin\theta} + A_\parallel \right],$$
$$g_2(x, Q^2) = \frac{y F_1(x, Q^2)}{2d} \left[ \frac{E + E'\cos\theta}{E'\sin\theta} A_\perp - A_\parallel \right], \tag{6}$$

where $\gamma = 2Mx/\sqrt{Q^2}$, $\theta$ is the scattering angle, $y = (E - E')/E$, $d = [(1-\epsilon)(2-y)]/[y(1+\epsilon R(x,Q^2))]$, and $\epsilon^{-1} = 1 + 2[1+\gamma^{-2}]\tan^2(\theta/2)$. For $F_1(x, Q^2) = F_2(x, Q^2)(1+\gamma^2)/[2x(1+R(x,Q^2))]$ we used the NMC [22] fits to $F_2$ data and the SLAC fit [23] to $R$, which was extrapolated to unmeasured regions for $x < 0.08$. All results were calculated using 28 $x$ bins for 4.5° and 20 $x$ bins for 7°. For tables and plots, every four bins were combined by error weighted averaging.

Results for $A_2$ for the proton and deuteron are shown in Fig. 1. and Tables 1-4. The error bars are statistical only. The systematic errors, dominated by radiative correction uncertainties, are indicated by bands. For a given $x$, the $Q^2$ probed by the two spectrometers differs by nearly a factor of two. The data agree within errors despite the differences in $Q^2$ of the measurements. Also in Fig. 1 are the proton results from SMC [4], and the $\sqrt{R}$ [23] positivity limits for each data set. The data are much closer to zero than the positivity limit, however, $A_2^p$ is consistently $> 0$. Since $A_2$ is expected to be zero at high $Q^2$ (because $R \to 0$), these data indicate that $A_2$ must have $Q^2$ dependence. A comparison of the data with the hypothesis $A_2 = 0$ yields $\chi^2 = 73$ for the proton and $\chi^2 = 44$ for the deuteron for 48 degrees of freedom (dof).

Measurements of $g_2$ for the proton and deuteron are shown in Tables 1-4 and Fig. 2 shows $xg_2$. The $g_2^d$ results are per nucleon. The systematic errors are indicated by bands. Also shown is the $g_2^{WW}$ curve evaluated using Eq. (2) at $E = 29$ GeV and $\theta = 4.5°$. The same curve for $\theta = 7°$ is nearly indistinguishable. The values for $g_2^{WW}$ were determined from $g_1(x, Q^2)$ evaluated from a fit to world data of $A_1$ [24] and assuming negligible higher-twist contributions. Also shown are the bag model predictions of Stratmann [25] and Song and McCarthy [17], which include both twist-2 and twist-3 contributions for $Q^2 = 5$ (GeV/c)$^2$. At high $x$ the results for $g_2^p$ indicate a negative trend consistent with the expectations for $g_2^{WW}$. A comparison of the proton data with the hypothesis $g_2 = 0$ yields a $\chi^2$ of 52 for 48 dof while a comparison with the hypothesis $g_2 = g_2^{WW}$ yields a $\chi^2$ of 43. The corresponding confidence levels for agreement with the hypotheses are 32% and 67%, respectively. The deuteron results are less conclusive because of the larger errors. The $\chi^2$ tests for $g_2 = 0$ and $g_2 = g_2^{WW}$ yield similar $\chi^2$ values of around 45 for 48 dof.

By extracting the quantity $\overline{g_2}(x, Q^2) = g_2(x, Q^2) - g_2^{WW}(x, Q^2)$, we can look for possible quark mass and higher twist effects. If the term in Eq. (1) which depends on quark masses can be neglected then $\overline{g_2}(x, Q^2)$ is entirely twist-3. Our results can be seen in Tables 1-4, and from the difference between the data and the solid line in Fig. 2. Within the experimental uncertainty the data are consistent with $\overline{g_2}$ being zero but also with $\overline{g_2}$ being of the same order of magnitude as $g_2^{WW}$.

Using our results for the longitudinal spin structure functions $g_1^p$ and $g_1^d$, we have computed the first few moments of the OPE sum rules, and solved for the twist-3 matrix elements $d_n$. These moments are defined to be $\Gamma_1^{(n)} = \int_0^1 x^n g_1(x) dx$ and $\Gamma_2^{(n)} = \int_0^1 x^n g_2(x) dx$. For the measured region $0.03 < x < 0.8$, we evaluated $g_1$ and corrected the twist-2 part of $g_2$ to fixed $Q^2 = 5$ (GeV/c)$^2$ assuming $g_1/F_1$ is independent of $Q^2$ [24], and have averaged the two spectrometer results to evaluate the moments. Possible $Q^2$ dependence of $\overline{g_2}$ has been neglected. We neglect the contribution from the region $0 \leq x < 0.03$ because of the $x^n$ suppression factor. For $0.8 < x \leq 1$, we assume that both $g_1$ and $g_2$ behave as $(1-x)^3$ since at high $x$, $g_2 \approx -g_1$, and we fit the data for $x > 0.56$. The uncertainty in the extrapolated contribution is taken to be the same as the contribution itself. The results are shown in Table 5a. We find that the $d_n$ results are somewhat sensitive to using the different assumption that $A_1$ and $A_2$ are independent of $Q^2$. However, we believe that the $A_2$ assumption is not valid given our nonzero results presented here. For comparison, in Table 5b we quote theoretical predictions [17,25–27] for $d_2^p$ and $d_2^d$. For $d_2^d$ the proton and neutron results were averaged and a deuteron D-state correction was applied. Our extracted values for $d_n$ are consistent with zero, but the errors are large. We note that the results for $d_2^p$ and $d_2^d$ differ in sign from the theoretical QCD sum rule calculations [26,27]. The bag model predictions [17,25], however, are of the same sign as the data. We have also evaluated the integrals $\int_{0.03}^1 g_2^p(x) dx = -0.013 \pm 0.028$ and $\int_{0.03}^1 g_2^d(x) dx = -0.033 \pm 0.082$ using the same high-$x$ extrapolation as discussed above. These results are also consistent with zero.

In summary, we have measured the proton and deuteron spin structure function $g_2$ and virtual photon-nucleon asymmetry $A_2$ as a function of $x$ at two different $Q^2$. We find that $A_2$ is significantly smaller than the $\sqrt{R}$ limit. We also find that $A_2 > 0$ for the proton. Within errors $g_2$ is consistent with the twist-2 $g_2^{WW}$ calculation, and is also in agreement with some theoretical predictions [17,25]. The component $\overline{g_2}$ is consistent with zero, but also with $\overline{g_2}$ being of the same order of magnitude as $g_2^{WW}$. Twist-3 matrix elements $d_n$ have been evaluated from the moments of $g_1$ and $g_2$. Within errors the results are consistent with zero. More precise data on $g_2$ are needed in order to make any conclusions regarding possible twist-3 and quark-mass-dependent contributions.

We wish to thank N. Shumeiko for help with the radiative corrections, and R. Jaffe and J. Ralston for useful discussions. This work was supported by Department of Energy contracts: DE-AC05-84ER40150, W-2705-Eng-48, DE-FG05-94ER40859, DE-AC03-76SF00515, DE-FG03-88ER40439, DE-FG05-88ER40390, DEFG05-86ER40261, and DE-AC02-76ER00881; by National Science Foundation Grants 9114958, 9307710, 9217979, 9104975, and 9118137; by the Schweizersche Nationalfonds; by the Commonwealth of Virginia; by the Centre National de la Recherche Scientifique and the Commissariat a l'Energie Atomique (French groups); and by the Japanese Ministry of Education, Science, and Culture.

# REFERENCES

∗ Permanent address: Oliver Lodge Lab, University of Liverpool, Liverpool, U. K.

∗∗ Permanent address: University of Bonn, D-53113 Bonn, Germany.

† Present address: College of William and Mary, Williamsburg, Virginia 23187.

§ Permanent address: FFIYM, P.O. Box 25, N-2007 Kjeller, Norway.

‡ Present address: Kent State University, Kent, Ohio 44242.

†† Permanent address: CERN, 1211 Geneva 23, Switzerland.

‡‡ Present Address: California Institute of Technology, Pasadena, CA 91125

[1] EMC, J. Ashman *et al.,* Phys. Lett. **B206**, 364 (1988); Nucl. Phys. **B328**, 1 (1989).

[2] SMC, D. Adams *et al.*, Phys. Lett. **B329**, 399 (1994).

[3] SMC, B. Adeva *et al.*, Phys. Lett. **B302**, 533 (1993).

[4] SMC, D. Adams *et al.*, Phys. Lett. **B336**, 125 (1994).

[5] SLAC E142, P. L. Anthony *et al.,* Phys. Rev. Lett. **71**, 959 (1993).

[6] SLAC E143, K. Abe *et al.*, Phys. Rev. Lett. **74**, 346 (1995).

[7] SLAC E143, K. Abe *et al.*, Phys. Rev. Lett. **75**, 25 (1995).

[8] J. Ellis and R. Jaffe, Phys. Rev. **D9**, 1444 (1974); **D10**, 1669 (1974) (Erratum).

[9] J. D. Bjorken, Phys. Rev. **148**, 1467 (1966); Phys. Rev. **D1**, 1376 (1970).

[10] R. L. Jaffe and X. Ji, Phys. Rev. **D43**, 724 (1991).

[11] E. V. Shuryak and A. I. Vainshtein, Nucl. Phys. **B201**, 141 (1982).

[12] L. Mankiewicz and Z. Rysak, Phys. Rev. **D43**, 733 (1991).

[13] X. Artru and M. Mekfi, Z. Phys. **C45**, 669, (1990).

[14] J. L. Cortes, B. Pire and J. P. Ralston, Z. Phys. **C55**, 409 (1992).

[15] S. Wandzura and F. Wilczek, Phys. Lett. **B72**, 195 (1977).

[16] R. L. Jaffe and X. Ji, Phys. Rev. Lett. **67**, 552 (1991).

[17] X. Song and J. S. McCarthy, Phys. Rev. **D49**, 3169 (1994), **D50**, 4718, (1994) (Erratum), X. Song, INPP-UVA-95/04 (1995).

[18] H. Burkhardt and W. N. Cottingham, Ann. Phys. **56**, 453 (1970).

[19] M. Anselmino, A. Efremov and E. Leader, preprint CERN-TH/7126/94, submitted to Phys. Rep.

[20] T. V. Kukhto and N. M. Shumeiko, Nucl. Phys. **B219**, 412 (1983); I. V. Akusevich and N. M. Shumeiko, J. Phys. **G20**, 513 (1994).

[21] Y. S. Tsai, Report No. SLAC-PUB-848 (1971); Y. S. Tsai, Rev. Mod. Phys. **46**, 815 (1974).

[22] NMC, P. Amaudruz *et al.,* Phys. Lett. **B295**, 159 (1992).

[23] L. W. Whitlow *et al.,* Phys. Lett. **B250**, 193 (1990).

[24] SLAC E143 K. Abe *et al.*, SLAC-PUB-95-6997 (1995), submitted to Phys. Lett. B.

[25] M. Stratmann, Z. Phys. **C60**, 763 (1993), and private communication for values at $Q^2 = 5$ $(\text{GeV}/\text{c})^2$.

[26] E. Stein *et al.*, Phys. Lett. **B334**, 369 (1995).

[27] I. I. Balitsky, V. M. Braun and A. V. Kolesnichenko, Phys. Lett. **B242**, 245 (1990); **B318**, 648 (1993) (Erratum).

Table 1: Results for $A_2$, $g_2$ and $\overline{g_2}$ for the proton measured in the 4.5° spectrometer at the indicated average values of $x$ and $Q^2$. The highest $x$ bin shown is in the resonance region defined by missing mass $W^2 < 4$ GeV$^2$.

| $x$ interval | $<x>$ | $<Q^2>$ (GeV/c)$^2$ | $A_2^p$ ±stat ±syst | $g_2^p$ ±stat ±syst | $\overline{g_2^p}$ ±stat ±syst |
|---|---|---|---|---|---|
| .029 − .047 | .038 | 1.49 | .016 ± .018 ± .006 | .492 ± .981 ± .287 | .219 ± .982 ± .294 |
| .047 − .075 | .060 | 2.01 | .024 ± .014 ± .005 | .403 ± .374 ± .128 | .209 ± .375 ± .136 |
| .075 − .120 | .095 | 2.60 | .004 ± .015 ± .005 | −.234 ± .203 ± .073 | −.303 ± .204 ± .081 |
| .120 − .193 | .152 | 3.21 | .020 ± .021 ± .008 | −.135 ± .125 ± .048 | −.135 ± .126 ± .054 |
| .193 − .310 | .242 | 3.77 | .090 ± .032 ± .011 | −.046 ± .079 ± .025 | .025 ± .080 ± .031 |
| .310 − .498 | .379 | 4.21 | .134 ± .060 ± .013 | −.051 ± .048 ± .009 | .052 ± .050 ± .014 |
| .498 − .799 | .595 | 4.55 | .044 ± .153 ± .021 | −.037 ± .020 ± .003 | −.005 ± .022 ± .005 |

Table 2: Results for $A_2$, $g_2$ and $\overline{g_2}$ for the proton measured in the 7.0° spectrometer at the indicated average values of $x$ and $Q^2$.

| $x$ interval | $<x>$ | $<Q^2>$ (GeV/c)$^2$ | $A_2^p$ ±stat ±syst | $g_2^p$ ±stat ±syst | $\overline{g_2^p}$ ±stat ±syst |
|---|---|---|---|---|---|
| .075 − .120 | .100 | 3.76 | .025 ± .025 ± .007 | .060 ± .367 ± .097 | −.002 ± .368 ± .104 |
| .120 − .193 | .154 | 4.97 | .048 ± .019 ± .007 | .172 ± .141 ± .047 | .197 ± .142 ± .054 |
| .193 − .310 | .243 | 6.36 | .049 ± .022 ± .007 | −.064 ± .070 ± .020 | .010 ± .071 ± .028 |
| .310 − .498 | .382 | 7.75 | .075 ± .035 ± .008 | −.039 ± .034 ± .006 | .043 ± .035 ± .011 |
| .498 − .799 | .585 | 8.85 | .102 ± .083 ± .013 | −.022 ± .011 ± .001 | −.000 ± .012 ± .003 |

Table 3: Results for $A_2$, $g_2$ and $\overline{g_2}$ for the deuteron measured in the 4.5° spectrometer at the indicated average values of $x$ and $Q^2$. The highest $x$ bin shown is in the resonance region defined by missing mass $W^2 < 4$ GeV$^2$.

| $x$ interval | $<x>$ | $<Q^2>$ (GeV/c)$^2$ | $A_2^d$ ±stat ±syst | $g_2^d$ ±stat ±syst | $\overline{g_2^d}$ ±stat ±syst |
|---|---|---|---|---|---|
| .029 − .047 | .038 | 1.49 | .070 ± .045 ± .008 | 3.428 ± 2.158 ± .376 | 3.278 ± 2.158 ± .378 |
| .047 − .075 | .060 | 2.01 | −.025 ± .028 ± .006 | −.658 ± .708 ± .153 | −.794 ± .708 ± .156 |
| .075 − .120 | .095 | 2.60 | .007 ± .032 ± .010 | .009 ± .391 ± .117 | −.050 ± .391 ± .118 |
| .120 − .193 | .152 | 3.21 | .004 ± .045 ± .015 | −.117 ± .244 ± .079 | −.102 ± .244 ± .080 |
| .193 − .310 | .242 | 3.77 | .079 ± .072 ± .019 | .127 ± .154 ± .040 | .134 ± .155 ± .042 |
| .310 − .498 | .379 | 4.22 | −.077 ± .145 ± .017 | −.128 ± .094 ± .010 | −.094 ± .095 ± .012 |
| .498 − .799 | .594 | 4.55 | .344 ± .390 ± .044 | .037 ± .039 ± .003 | .030 ± .040 ± .004 |

Table 4: Results for $A_2$, $g_2$ and $\overline{g_2}$ for the deuteron measured in the 7.0° spectrometer at the indicated average values of $x$ and $Q^2$.

| $x$ interval | $<x>$ | $<Q^2>$ (GeV/c)$^2$ | $A_2^d$ ±stat ±syst | $g_2^d$ ±stat ±syst | $\overline{g_2^d}$ ±stat ±syst |
|---|---|---|---|---|---|
| .075 − .120 | .100 | 3.76 | .025 ± .046 ± .010 | .167 ± .622 ± .126 | .094 ± .622 ± .128 |
| .120 − .193 | .154 | 4.97 | −.007 ± .036 ± .011 | −.107 ± .236 ± .073 | −.092 ± .236 ± .074 |
| .193 − .310 | .242 | 6.36 | −.042 ± .043 ± .015 | −.134 ± .117 ± .038 | −.114 ± .117 ± .040 |
| .310 − .498 | .382 | 7.75 | −.002 ± .073 ± .014 | −.042 ± .056 ± .010 | −.007 ± .057 ± .011 |
| .498 − .799 | .584 | 8.84 | .217 ± .183 ± .028 | .000 ± .018 ± .002 | .008 ± .019 ± .002 |

| Table 5a: Results for the moments $\Gamma_1^{(n)}$ and $\Gamma_2^{(n)}$ evaluated at $Q^2 = 5$ (GeV/c)$^2$, and the extracted twist-3 matrix elements $d_n$ for proton (p) and deuteron (d) targets. The errors include statistical (which dominate) and systematic contributions. | | | | |
|---|---|---|---|---|
| | n | $\Gamma_1^{(n)} \times 10^3$ | $\Gamma_2^{(n)} \times 10^3$ | $d_n \times 10^3$ |
| p | 2 | $12.1 \pm 1.0$ | $-6.3 \pm 1.8$ | $5.4 \pm 5.0$ |
| | 4 | $3.2 \pm 0.4$ | $-2.3 \pm 0.6$ | $0.7 \pm 1.7$ |
| | 6 | $1.2 \pm 0.2$ | $-1.0 \pm 0.3$ | $0.1 \pm 0.8$ |
| d | 2 | $4.0 \pm 0.8$ | $-1.4 \pm 3.0$ | $3.9 \pm 9.2$ |
| | 4 | $0.8 \pm 0.3$ | $0.0 \pm 1.0$ | $1.7 \pm 2.6$ |
| | 6 | $0.2 \pm 0.2$ | $0.1 \pm 0.5$ | $0.6 \pm 1.1$ |

| Table 5b: Theoretical predictions for the twist-3 matrix element $d_2^p$ for proton and $d_2^d$ for deuteron. | | | | |
|---|---|---|---|---|
| | Bag models | | QCD sum rules | |
| | Ref. [17] | Ref. [25] | Ref. [26] | Ref. [27] |
| $Q^2$ (GeV/c)$^2$ | 5 | 5 | 1 | 1 |
| $d_2^p \times 10^3$ | 17.6 | 6.0 | $-6 \pm 3$ | $-3 \pm 3$ |
| $d_2^d \times 10^3$ | 6.6 | 2.9 | $-17 \pm 5$ | $-13 \pm 5$ |

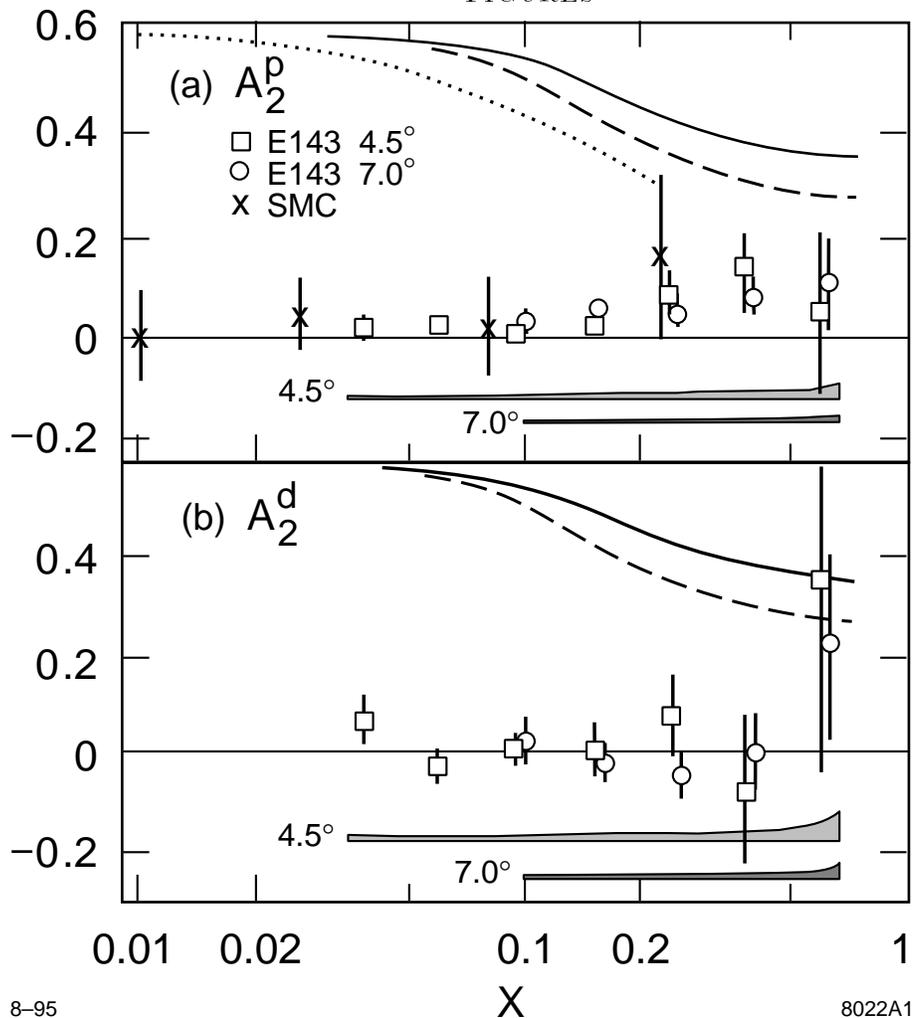

Fig. 1. Asymmetry measurements for (a) $A_2^p$, and (b) $A_2^d$ from E143 (two data sets) and SMC as a function of $x$. Systematic errors are indicated by bands. The curves show the $\sqrt{R}$ positivity constraints for the three data sets as determined by the SLAC parametrization [23] of $R$. The solid, dashed and dotted curves correspond to the 4.5° E143, 7.0° E143, and SMC kinematics, respectively. The symbols in (b) are as in (a). Overlapping data have been shifted slightly in $x$ to make errors clearly visible.

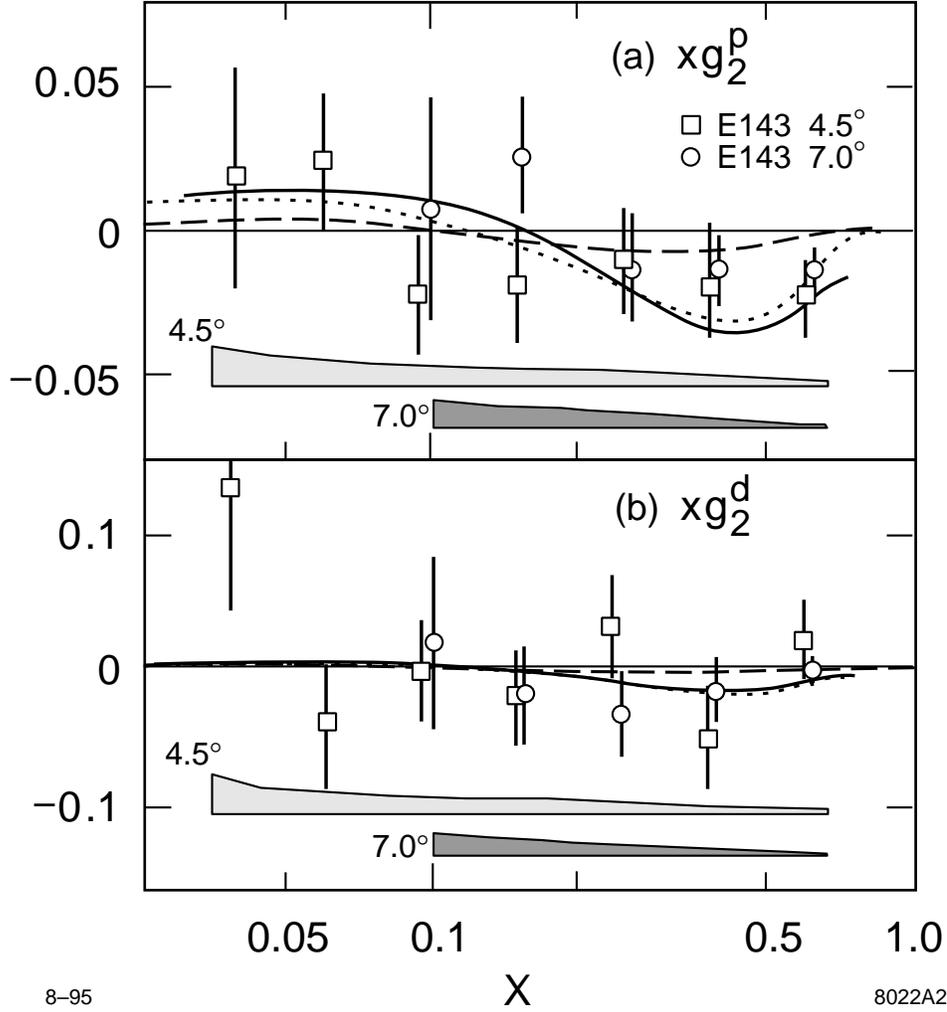

Fig. 2. Spin structure function measurements for (a) $xg_2^p$, and (b) $xg_2^d$ from this experiment (E143) as a function of $x$. Systematic errors are indicated by bands. The symbols in (b) are as in (a). Overlapping data have been shifted slightly in $x$ to make errors clearly visible. The solid curve shows the twist-2 $g_2^{WW}$ calculation for the kinematics of the 4.5° spectrometer. The same curve for 7° is nearly indistinguishable. The bag model calculations at $Q^2 = 5.0$ (GeV/c)$^2$ by Stratmann [25] (dotted) and Song and McCarthy [17] (dashed) are indicated.